\documentclass[a4paper]{jpconf}
\usepackage{graphicx}

\makeatletter
\def\slash#1{{\mathpalette\c@ncel{#1}}} 
\makeatother

\newcommand\beq{\begin{eqnarray}}
\newcommand\eeq{\end{eqnarray}}

\newcommand\la{\langle}
\newcommand\ra{\rangle}

\begin{document}
\title{Probing the twist-3 multi-gluon correlation functions by
$p^{\uparrow}p\to DX$}

\author{Yuji Koike${}^*$ and Shinsuke Yoshida${}^{\dagger}$}

\address{${}^*$Department of Physics, Niigata University, Ikarashi, Niigata 950-2181, Japan}
\address{${}^{\dagger}$Graduate School of Science and Technology, Niigata University, Ikarashi, Niigata 950-2181, Japan}

\ead{yoshida@nt.sc.niigata-u.ac.jp}

\begin{abstract}
We study the single spin asymmetry (SSA) for the $D$-meson 
production $A_N^D$ in the $pp$ collision, $p^\uparrow p\to DX$, in the framework of the 
collinear factorization.  Since the charm quark is mainly produced through
the $c\bar{c}$-pair creation from the gluon-fusion process, this is an ideal process
to probe the twist-3 triple-gluon correlation functions in the polarized nucleon. 
We derive the corresponding cross section 
formula for the contribution of the triple-gluon correlation function
to $A_N^D$ in $p^\uparrow p\to DX$, 
applying the method developed for $ep^\uparrow\to eDX$ in our previous study.  
As in the case of $ep^\uparrow\to eDX$, our result differs from a previous 
study in the literature.
We will also present a simple estimate of the triple-gluon correlation functions
based on the preliminary data on $A_N^D$ by RHIC. 
\end{abstract}

\section{Introduction}
\vspace{1mm}

The single spin asymmetry (SSA) in inclusive hard processes appears as a twist-3
observable in the collinear factorization which is valid in describing the large-$P_T$
hadron productions.  In this twist-3 mechanism, SSA is represented in terms 
of multi-parton correlations 
in the hadrons.   Such correlations in the transversely polarized nucleon
can be represented by the 
quark-gluon correlation functions and the triple-gluon correlation functions.  
So far the former effect has been investigated for various processes.  
To probe the latter effect,  
open-charm production in $ep$ and $pp$ collisions is an ideal tool, since the
$c\bar{c}$ pair is created mainly by the gluon-photon or gluon-gluon fusion processes.  

In our recent paper\,\cite{BKTY}, we formulated a method of 
calculating the contribution of the triple-gluon correlation functions 
to the single-spin-dependent cross section for
the $D$-meson production in semi-inclusive deep inelastic scattering (SIDIS),
$ep^\uparrow \to eDX$.  There we identified the complete set of the
relevant triple-gluon correlation functions and derived the corresponding 
cross section in the leading order with respect to the QCD coupling constant.  
Our result differed from the previous study in the literature\,\cite{KQ}, and 
we have clarified the origin of the discrepancy.  

In this work, we will apply the formalism of \cite{BKTY} to the
$pp$-collision, $p^\uparrow p\to DX$, and derive the corresponding single-spin-dependent cross section.  
This study is relevant to the ongoing RHIC experiment.  We also
provide a simple estimate for the constraint on the triple-gluon correlation functions, using a
preliminary data on $p^\uparrow p\to DX$ reported at RHIC\,\cite{spin2008}.

\section{Triple-gluon correlation functions for the transversely polarized nucleon}
\vspace{1mm}

Triple-gluon correlation functions in the transversely 
polarized nucleon are defined as the correlation functions of the
three gluon's field strength tensor $F^{\alpha\beta}$.   
Since there are two ways of constructing color-singlet correlation functions from the contraction with 
the symmetric and antisymmetric structure constants for the color SU(3) group, 
$d_{bca}$ and $f_{bca}$, one can define two gauge-invariant correlation functions in the nucleon as\,\cite{BKTY}
\beq
&&\hspace{-0.8cm}O^{\alpha\beta\gamma}(x_1,x_2)
=-g(i)^3\int{d\lambda\over 2\pi}\int{d\mu\over 2\pi}e^{i\lambda x_1}
e^{i\mu(x_2-x_1)}\la pS|d_{bca}F_b^{\beta n}(0)F_c^{\gamma n}(\mu n)F_a^{\alpha n}(\lambda n)
|pS\ra \nonumber\\
&&=2iM_N\left[
O(x_1,x_2)g^{\alpha\beta}\epsilon^{\gamma pnS}
+O(x_2,x_2-x_1)g^{\beta\gamma}\epsilon^{\alpha pnS}
+O(x_1,x_1-x_2)g^{\gamma\alpha}\epsilon^{\beta pnS}\right]
\label{3gluonO},\\
&&\hspace{-0.8cm}N^{\alpha\beta\gamma}(x_1,x_2)
=-g(i)^3\int{d\lambda\over 2\pi}\int{d\mu\over 2\pi}e^{i\lambda x_1}
e^{i\mu(x_2-x_1)}\la pS|if_{bca}F_b^{\beta n}(0)F_c^{\gamma n}(\mu n)F_a^{\alpha n}(\lambda n)
|pS\ra \nonumber\\
&&=2iM_N\left[
N(x_1,x_2)g^{\alpha\beta}\epsilon^{\gamma pnS}
-N(x_2,x_2-x_1)g^{\beta\gamma}\epsilon^{\alpha pnS}
-N(x_1,x_1-x_2)g^{\gamma\alpha}\epsilon^{\beta pnS}\right],  
\label{3gluonN}
\eeq
where $M_N$ is the nucleon mass, $S$ is the transverse-spin vector 
for the nucleon, 
$n$ is the light-like vector satisfying $p\cdot n=1$ and
we used the shorthand notation as $F^{\beta n}\equiv F^{\beta\rho}n_{\rho}$ {\it etc}.  
Hermiticity, $PT$-invariance and permutation
symmetry lead to
the decomposition of (\ref{3gluonO}) and (\ref{3gluonN}) in terms of the 
two real functions $O(x_1,x_2)$ and
$N(x_1,x_2)$ which have the following symmetry properties, 
\beq
&&O(x_1,x_2)=O(x_2,x_1),\hspace{1cm}O(x_1,x_2)=O(-x_1,-x_2), \nonumber\\
&&N(x_1,x_2)=N(x_2,x_1),\hspace{1cm}N(x_1,x_2)=-N(-x_1,-x_2).
\eeq
The gauge-link operator which restores gauge invariance of the correlation functions
is suppressed in (\ref{3gluonO}) and (\ref{3gluonN}) for simplicity.  

\section{Polarized cross section formula for $p^{\uparrow}p\to DX$}
\vspace{1mm}

The formalism for calculating the contribution of the triple-gluon correlation functions 
to $ep^\uparrow\to eDX$ developed in \cite{BKTY} can be directly applied to $p^\uparrow p\to DX$\,\cite{KY}.  
In this process, a $c\bar{c}$-pair is created by the gluon-fusion process, and 
the initial-state-interaction (ISI) 
diagrams as well as the final-state-interaction (FSI) diagrams 
give rise to the single-spin-dependent cross section as a pole contribution at $x_1=x_2$.  
The twist-3 cross section for
$p^{\uparrow}(p,S_\perp) + p(p') \to D(P_h)  + X$ with the center-of-mass energy $\sqrt{s}$
is given by 
\beq
&&\hspace{-0.7cm}
 P_{h}^0\frac{d\sigma^{\rm 3-gluon}}{d^3P_h}=
\frac{\alpha_s^2M_N\pi}{s}\sum_{f=c\bar{c}}\int\frac{dx'}{x'}G(x')\int\frac{dz}{z^2}D_f(z)\int\frac{dx}{x}\delta
 (\tilde{s}+\tilde{t}+\tilde{u})\epsilon^{p_c p n S_{\perp}}{1\over \tilde{u}}
 \nonumber\\
&&\hspace{-0.7cm}
\times\biggl[\delta_f\Bigl\{
\Bigl(\frac{d}{dx}O(x,x)-\frac{2O(x,x)}{x}\Bigr)\hat{\sigma}^{O1}
+\Bigl(\frac{d}{dx}O(x,0)-\frac{2O(x,0)}{x}\Bigr)\hat{\sigma}^{O2}
+\frac{O(x,x)}{x}\hat{\sigma}^{O3}
+\frac{O(x,0)}{x}\hat{\sigma}^{O4}
\Bigr\} \nonumber\\
&&\hspace{-0.7cm}
+\Bigl\{
\Bigl(\frac{d}{dx}N(x,x)-\frac{2N(x,x)}{x}\Bigr)\hat{\sigma}^{N1}
+\Bigl(\frac{d}{dx}N(x,0)-\frac{2N(x,0)}{x}\Bigr)\hat{\sigma}^{N2}
+\frac{N(x,x)}{x}\hat{\sigma}^{N3}
+\frac{N(x,0)}{x}\hat{\sigma}^{N4}
\Bigr\}
\biggr],\nonumber\\[2pt]
\label{result}
\eeq
where $\delta_c=1$ and $\delta_{\bar{c}}=-1$, $D_f(z)$ represents the 
$c\to D$ or $\bar{c}\to\bar{D}$ fragmentation functions, $G(x')$ is the unpolarized gluon density,
and $p_c$ is the four-momentum of the $c$ (or $\bar{c}$) quark (mass $m_c$) fragmenting into the
final $D$ (or $\bar{D}$) meson.  
The partonic hard cross sections in (\ref{result}) 
are given by  
\beq
\left\{
\begin{array}{lll}
 \hat{\sigma}^{O1}&=&
\left(\frac{1}{C_F}\frac{\tilde{u}-\tilde{t}}{\tilde{s}\tilde{t}\tilde{u}}
+\frac{1}{C_F}\frac{\tilde{u}}{\tilde{s}\tilde{t}^2}-\frac{1}{N^2C_F}\frac{\tilde{s}}{\tilde{t}^2\tilde{u}}\right)
\left(\tilde{t}^2+\tilde{u}^2+4m_c^2\tilde{s}-\frac{4m_c^4\tilde{s}^2}{\tilde{t}\tilde{u}}\right), \\[7pt]
 \hat{\sigma}^{O2}&=&
\left(\frac{1}{C_F}\frac{\tilde{u}-\tilde{t}}{\tilde{s}\tilde{t}\tilde{u}}
+\frac{1}{C_F}\frac{\tilde{u}}{\tilde{s}\tilde{t}^2}-\frac{1}{N^2C_F}\frac{\tilde{s}}{\tilde{t}^2\tilde{u}}
\right)\left(\tilde{t}^2+\tilde{u}^2+8m_c^2\tilde{s}-\frac{8m_c^4\tilde{s}^2}{\tilde{t}\tilde{u}}\right), \\[7pt]
 \hat{\sigma}^{O3}&=&
\left(\frac{1}{C_F}\frac{\tilde{u}-\tilde{t}}{\tilde{t}^2\tilde{u}^2}+\frac{1}{C_F}\frac{1}{\tilde{t}^3}-\frac{1}{N^2C_F}\frac{\tilde{s}^2}{\tilde{t}^3\tilde{u}^2}\right)
\left(8m_c^4\tilde{s}-4m_c^2\tilde{t}\tilde{u}\right), \\[7pt]
 \hat{\sigma}^{O4}&=&
\left(\frac{1}{C_F}\frac{\tilde{u}-\tilde{t}}{\tilde{t}^2\tilde{u}^2}+\frac{1}{C_F}\frac{1}{\tilde{t}^3}-\frac{1}{N^2C_F}\frac{\tilde{s}^2}{\tilde{t}^3\tilde{u}^2}\right)
\left(16m_c^4\tilde{s}-4m_c^2\tilde{t}\tilde{u}\right), \\
\end{array}
\right.
\label{hardO}
\eeq
\beq
\left\{
\begin{array}{lll}
 \hat{\sigma}^{N1}&=&
\left(\frac{1}{C_F}\frac{\tilde{t}^2+\tilde{u}^2}{\tilde{s}^2\tilde{t}\tilde{u}}
+\frac{1}{C_F}\frac{\tilde{u}}{\tilde{s}\tilde{t}^2}-\frac{1}{N^2C_F}\frac{\tilde{s}}{\tilde{t}^2\tilde{u}}\right)
\left(\tilde{t}^2+\tilde{u}^2+4m_c^2\tilde{s}-\frac{4m_c^4\tilde{s}^2}{\tilde{t}\tilde{u}}\right), \\[7pt]
 \hat{\sigma}^{N2}&=&
-\left(\frac{1}{C_F}\frac{\tilde{t}^2+\tilde{u}^2}{\tilde{s}^2\tilde{t}\tilde{u}}
+\frac{1}{C_F}\frac{\tilde{u}}{\tilde{s}\tilde{t}^2}-\frac{1}{N^2C_F}\frac{\tilde{s}}{\tilde{t}^2\tilde{u}}\right)
\left(\tilde{t}^2+\tilde{u}^2+8m_c^2\tilde{s}-\frac{8m_c^4\tilde{s}^2}{\tilde{t}\tilde{u}}\right), \\[7pt]
 \hat{\sigma}^{N3}&=&
\left(\frac{1}{C_F}\frac{\tilde{t}^2+\tilde{u}^2}{\tilde{s}\tilde{t}^2\tilde{u}^2}
+\frac{1}{C_F}\frac{1}{\tilde{t}^3}-\frac{1}{N^2C_F}\frac{\tilde{s}^2}{\tilde{t}^3\tilde{u}^2}\right)
\left(8m_c^4\tilde{s}-4m_c^2\tilde{t}\tilde{u}\right), \\[7pt]
 \hat{\sigma}^{N4}&=&
-\left(\frac{1}{C_F}\frac{\tilde{t}^2+\tilde{u}^2}{\tilde{s}\tilde{t}^2\tilde{u}^2}
+\frac{1}{C_F}\frac{1}{\tilde{t}^3}-\frac{1}{N^2C_F}\frac{\tilde{s}^2}{\tilde{t}^3\tilde{u}^2}\right)
\left(16m_c^4\tilde{s}-4m_c^2\tilde{t}\tilde{u}\right), \\
\end{array}
\right.
\label{hardN}
\eeq
where $N=3$ and $C_F=(N^2-1)/(2N)$, and 
$\tilde{s}$, $\tilde{t}$, $\tilde{u}$ are defined as
\beq
\tilde{s}=(xp+x'p')^2,\qquad\tilde{t}=(xp-p_c)^2-m_c^2,\qquad\tilde{u}=(x'p'-p_c)^2-m_c^2.
\eeq
The hard cross sections associated with the first term in the first parentheses in
(\ref{hardO}) and (\ref{hardN}) come from ISI, and those associated
with the second and the third terms in the same parentheses come from FSI.  
As in the case of $ep^\uparrow\to eDX$, the cross section in (\ref{result}) receives the contribution from the
four functions $O(x,x)$, $O(x,0)$, $N(x,x)$, $N(x,0)$. 
Unlike the case of SIDIS, presence of ISI gives rise to the different hard cross sections
for $O$ and $N$ functions.  
From (\ref{result}), it is clear that the process $p^\uparrow p\to DX$ itself 
is not sufficient for the complete separation of the four functions.
For the separation, the process $ep^\uparrow\to eDX$ serves greatly, since it has five
structure functions with different dependences on the azimuthal angles to which 
the four functions contribute differently\,\cite{BKTY}.

Our result in (\ref{result}) differs from a previous work\,\cite{KQVY}:  
The result in \cite{KQVY} is obtained from (\ref{result}) by omitting the terms
with $O(x,0)$ and $N(x,0)$ (and their derivatives) 
and by the replacement $O(x,x)\to O(x,x)+O(x,0)$ and $N(x,x)\to N(x,x)-N(x,0)$.  
This difference originates from an ad-hoc assumption in the factorization formula
in \cite{KQVY,KQ}.  We emphasize the appearance of the four different contributions with   
$\{O(x,x), O(x,0), N(x,x), N(x,0)\}$ is a consequence of the symmetry property
implied in the decomposition (\ref{3gluonO}) and (\ref{3gluonN}), in particular, 
the different coefficient tensors in front of $O(x,x)$ and $O(x,0)$ (likewise for $N(x,x)$ and $N(x,0)$)
at $x_1=x_2=x$ 
lead to different hard cross sections for the above four functions.   
See \cite{BKTY,KY} for more details.

\section{Numerical estimate}
\vspace{1mm}

As is shown in (\ref{result}), four nonperturbative functions 
are involved in the twist-3 cross section for $A_N^D$.  
At present there is no information on these functions.  
Preliminary data on $A_N^D$ by RHIC-PHENIX\,\cite{spin2008} suggests $|A_N^D|\leq 5$ \%.  
So we will present a model estimate on the upper bound of
these nonperturbative functions for several cases
by requiring that the calculated $A_N^D$ be less than 5 \%.  
We first found from (\ref{hardO}) and (\ref{hardN}) the relations 
$ |\hat{\sigma}^{O1,O2,N1,N2}| \gg |\hat{\sigma}^{O3,O4,N4,N4}|$ and
$\hat{\sigma}^{O1}\simeq \hat{\sigma}^{O2} \simeq \hat{\sigma}^{N1} \simeq -\hat{\sigma}^{N2}$   
for the RHIC kinematics (see below).  
Accordingly, if $O(x,x)$ and $O(x,0)$ have the same (opposite) sign, they contribute to $A_N^D$
constructively (destructively).  Likewise for $N(x,x)$ and $N(x,0)$.  
From this fact we present a model calculation of $A_N^D$
for two extreme cases:   
\beq
&&{\rm Model\ 1}:\qquad O(x,x)=-O(x,0)=N(x,x)=N(x,0), 
\label{case1}\\
&&{\rm Model\ 2}:\qquad O(x,x)=O(x,0)=N(x,x)=-N(x,0).
\label{case2}
\eeq
In Model 1 the contributions to $A_N^D$ from four functions become minimum, while
in Model 2 they become maximum.  Therefore larger magnitude for the functions
is allowed for Model 1 in order to make $|A_N^D|\leq 5$ \%. 
For the actual form of the nonperturbative functions, we follow \cite{KQ}
and set 
\beq
O(x,x)=K_GxG(x),
\label{gluon}
\eeq
where $G(x)$ is the unpolarized gluon distribution and 
$K_G$ is a constant which we determine so as to achieve $|A_N^D|\leq 5$ \% for the RHIC kinematics. 
Obviously, the ansatz (\ref{gluon}) is a very crude approximation and the result below should be taken 
only as an estimate of the order-of-magnitude.  
For the numerical calculation, we use GJR08 distribution\,\cite{GJR} for $G(x)$ and 
KKKS08 fragmentation function\,\cite{KKKS} for $D_f(z)$.  
We also assumed the same scale dependence for $O(x,x)$ {\it etc} as $G(x)$
for simplicity.  
We calculated $A_N$ for the $D$ and $\bar{D}$ mesons at the RHIC energy of $\sqrt{s}=200$ GeV and $P_T=2$ GeV
with the parameter $m_c=1.3$ GeV by setting the 
scale of all the distribution and fragmentation functions at $\mu=\sqrt{P_T^2+m_c^2}$.

Fig. 1 shows the result of $A_N^D$ for Model 1 with $K_G=0.005$.  
In order to see the behavior of each term
in (\ref{result}), we showed in the left figure of Fig. 1
the contribution to $A_N^{D^0}$ from the four functions
$O(x,x)$, $O(x,0)$, $N(x,x)$ and $N(x,0)$.  (For $N(x,0)$, $-A_N^D$ is shown.)
As is seen from this figure, if one sets all four functions identical,
contributions to $A_N^{D^0}$ from them are very close.  
In the middle and right figures of Fig. 1, $A_N$ for the $D^0$ and $\bar{D^0}$ mesons
are shown, respectively.  Even though each of four contribution gives rise to as large as 50 \% asymmetry
as shown in the left figure,
the resulting total $A_N$ is of $O(5\,\%)$ due to the cancellation among them.  
Therefore if the relation (\ref{case1}) approximately holds, $K_G=0.005$ in (\ref{gluon})
provides an upper limit for $O(x,x)$ {\it etc}. 
Because of the relation (\ref{case1}), the $O$-term and the $N$-term
contribute to $A_N$ constructively (destructively) for $D^0$ ($\bar{D^0}$) meson
as shown in Fig. 1.  This feature was also observed in \cite{KQVY}.

\begin{figure}[h]
\begin{center}

\scalebox{0.49}{\includegraphics{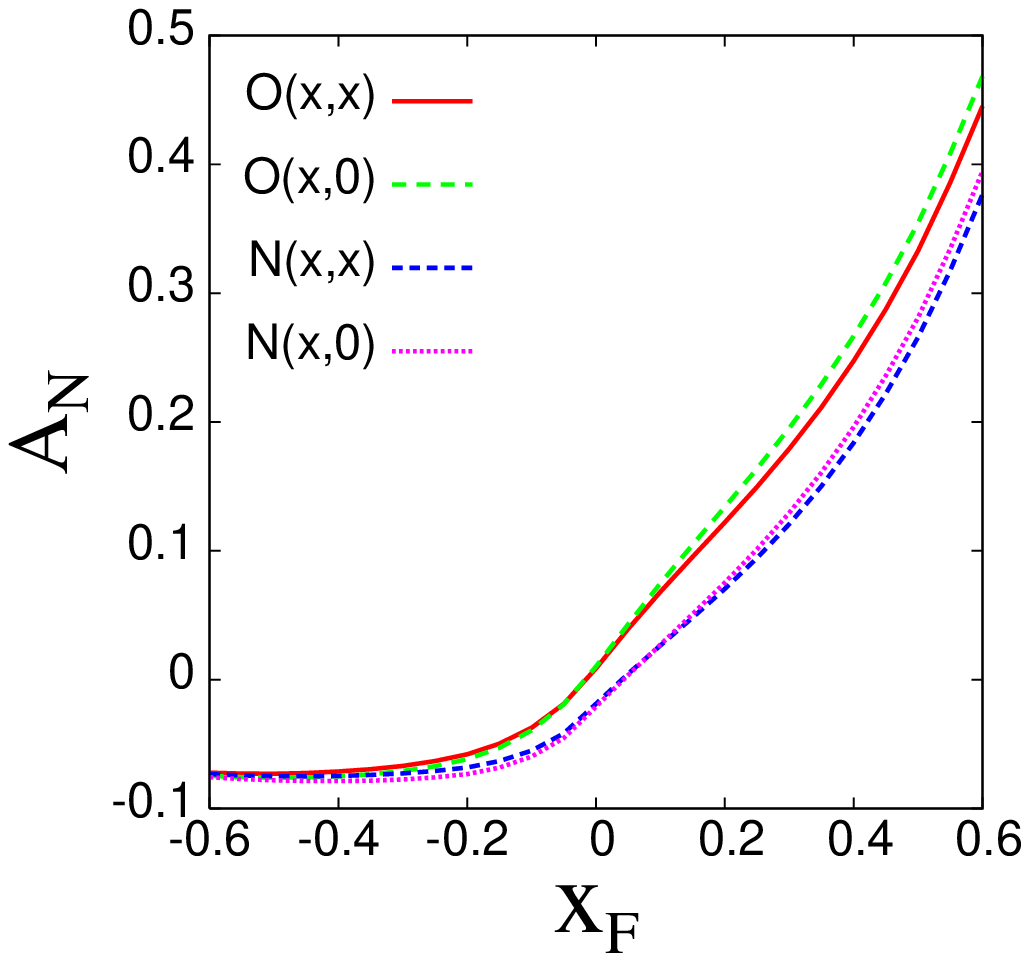}}\hspace{-5mm}
\scalebox{0.49}{\includegraphics{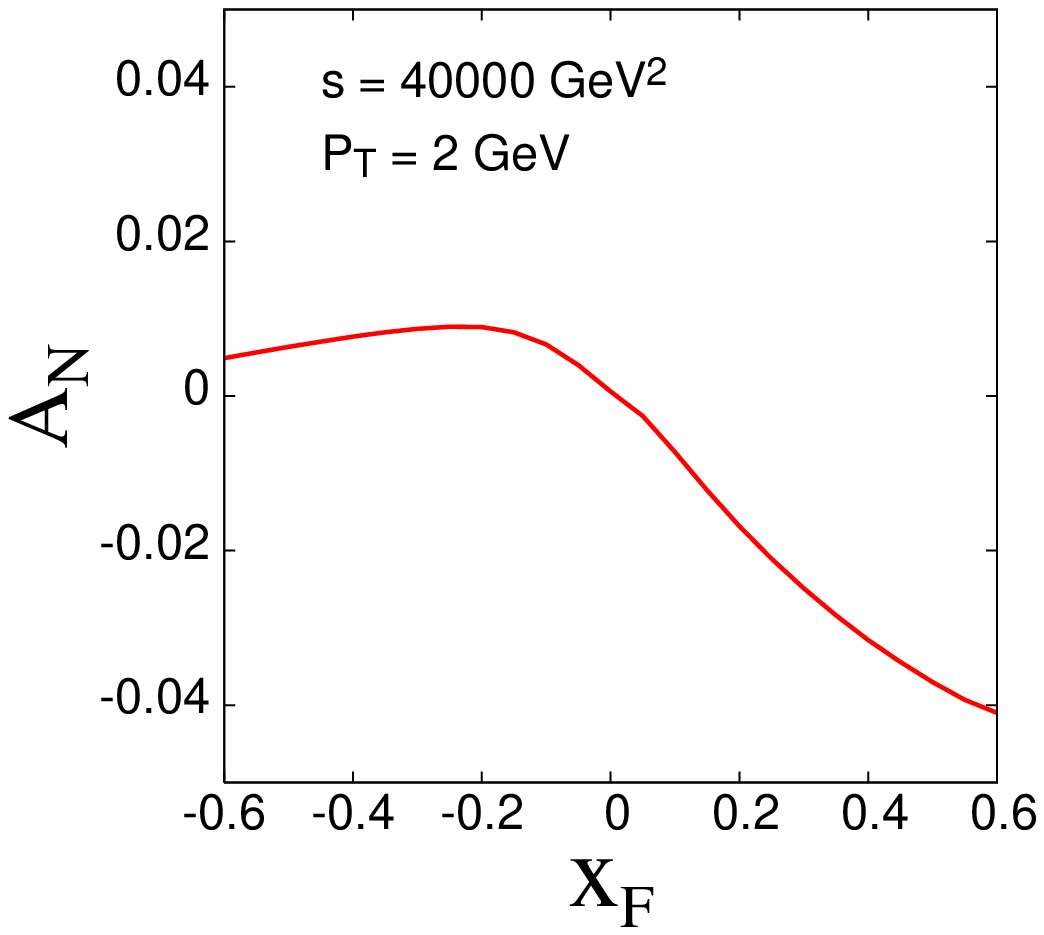}}\hspace{-5mm}
\scalebox{0.49}{\includegraphics{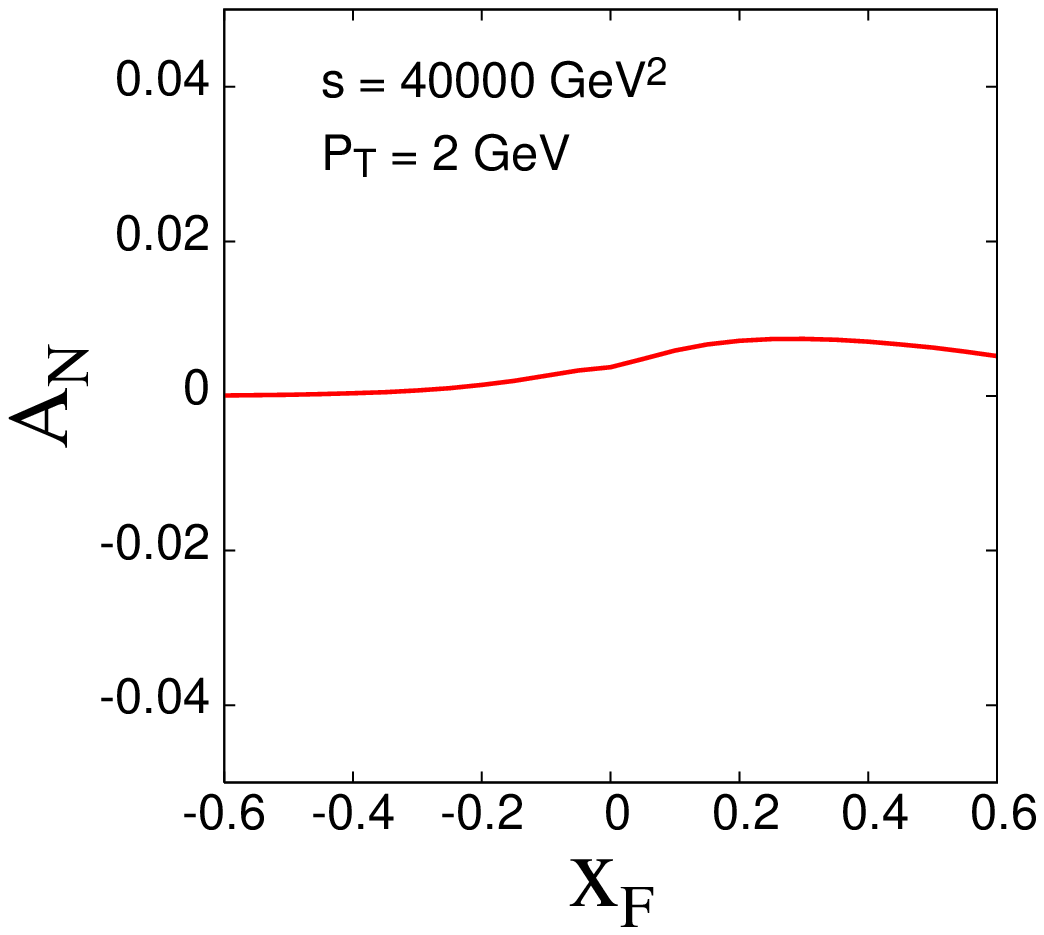}}

\caption{$A_N^D$ for Model 1.  Decomposition of $A_N^D$ into the 4 components (left).
$A_N$ for $D^0$ meson (middle).  $A_N$ for $\bar{D^0}$ meson (right).  }
\label{fig.case1}
\end{center}
\end{figure}

Fig. 2 shows the result for Model 2 with $K_G=0.0001$.  In the left figure of Fig. 2, the behavior of
each contribution is shown.  In the middle and right figures, 
$A_N$ for the $D^0$ and $\bar{D}^0$ mesons are shown, respectively. 
In this Model 2, we had to take $K_G$ as small as  $K_G\sim 0.0001$, since $O(x,x)$ and $O(x,0)$
as well as $N(x,x)$ and $N(x,0)$ contribute to $A_N$ of either $D^0$ or $\bar{D}^0$ constructively.  

\begin{figure}[h]
\begin{center}

\scalebox{0.49}{\includegraphics{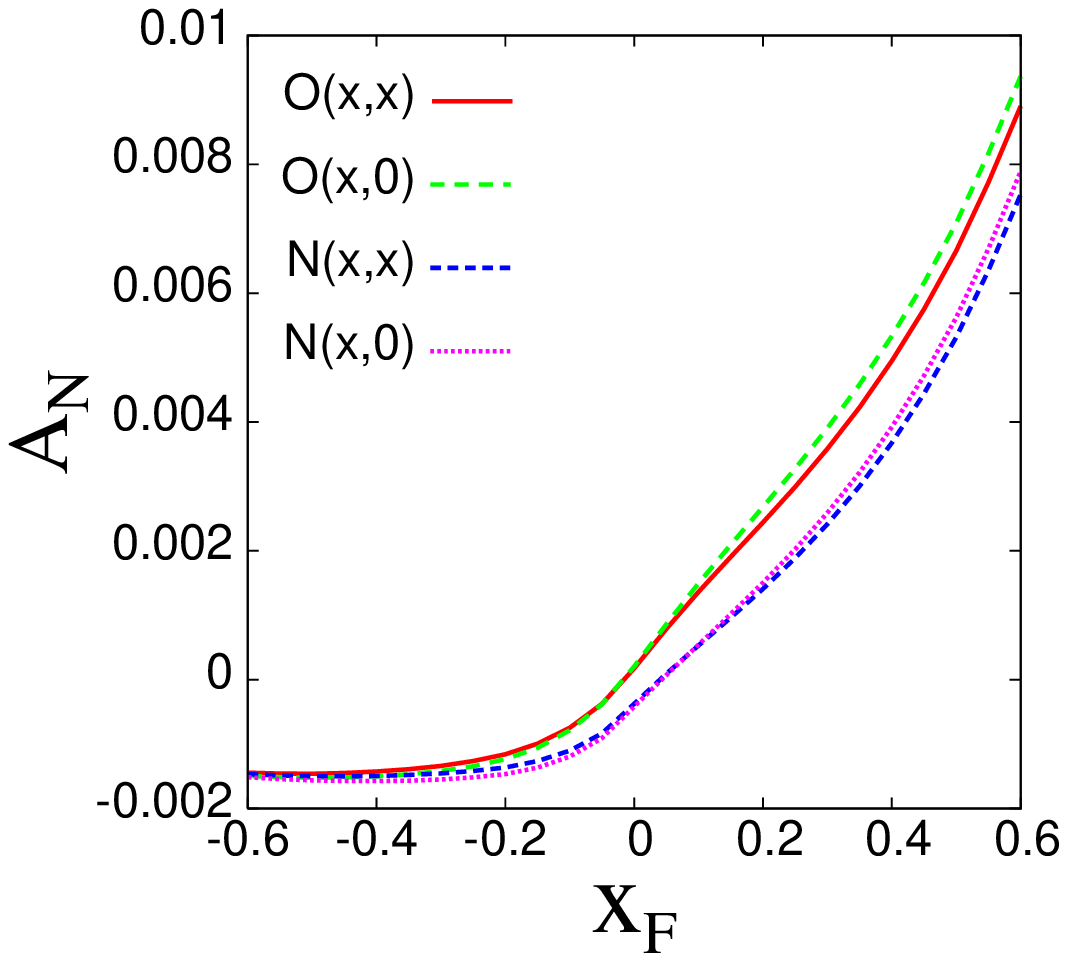}}\hspace{-5mm}
\scalebox{0.49}{\includegraphics{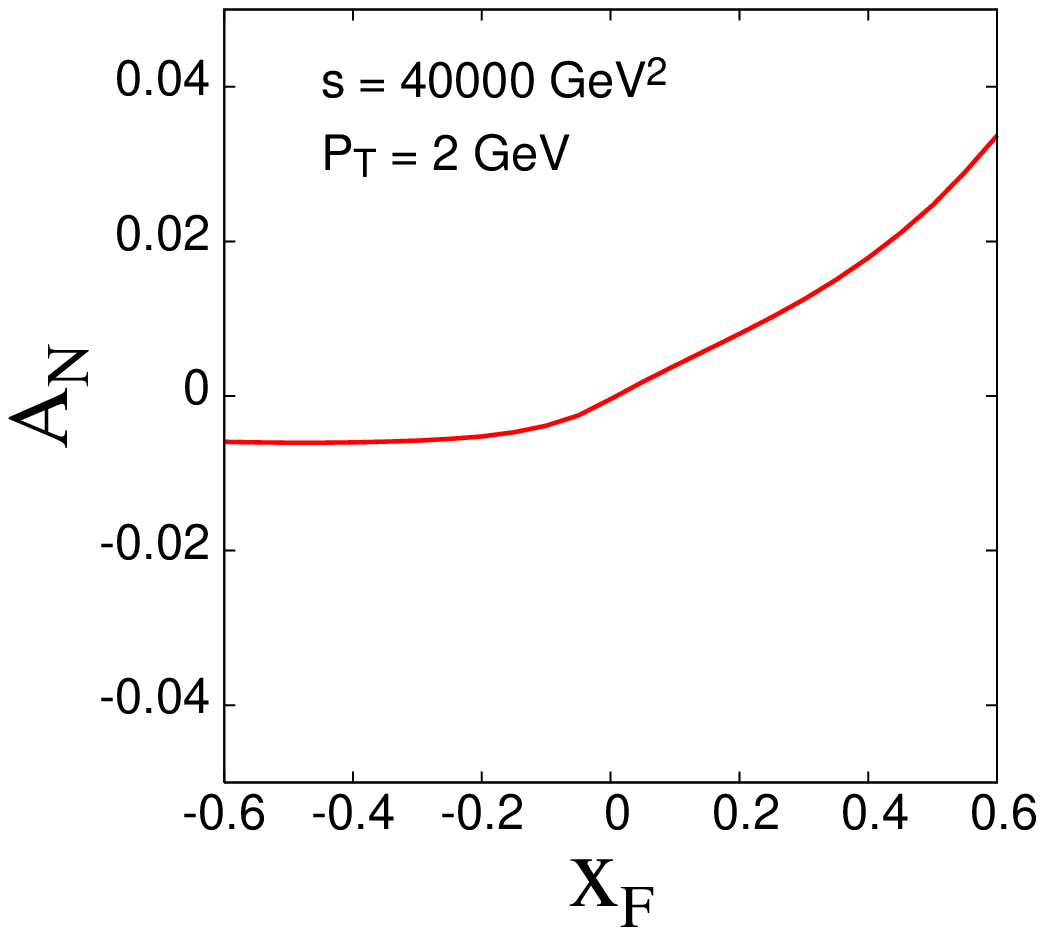}}\hspace{-5mm}
\scalebox{0.49}{\includegraphics{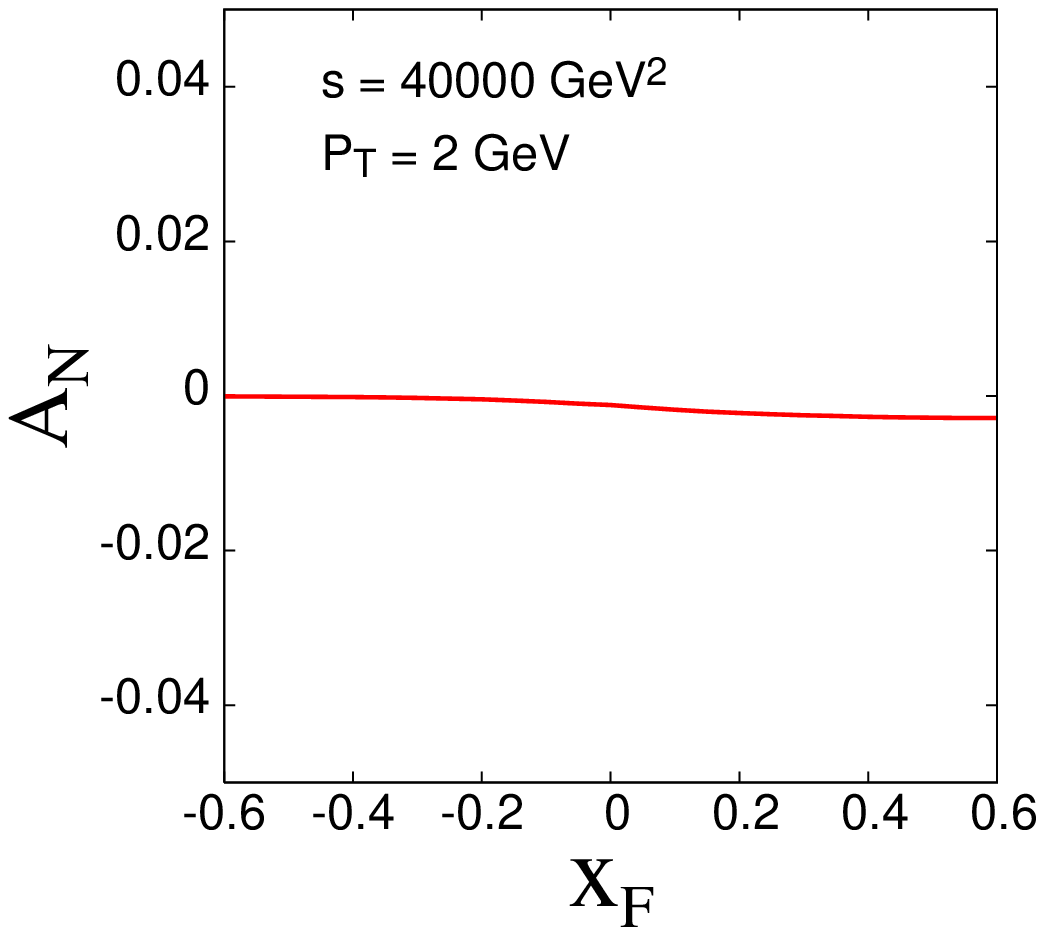}}

\caption{$A_N^D$ for Model 2.  Decomposition of $A_N^D$ into the 4 components (left).
$A_N$ for $D^0$ meson (middle).  $A_N$ for $\bar{D^0}$ meson (right).  }
\label{fig.case2}
\end{center}
\end{figure}

From the left figures of Figs. 1 and 2, one can see that the magnitude
of each of the four contributions is very similar if the four functions have the same
magnitude. 
This means that 
the contributions to $A_N^D$
from $O(x,x)+O(x,0)$ and $N(x,x)-N(x,0)$ are, respectively, much larger than
those from $O(x,x)-O(x,0)$ and $N(x,x)+N(x,0)$, because
\beq
|\hat{\sigma}^{O1}+\hat{\sigma}^{O2}| \gg |\hat{\sigma}^{O1}-\hat{\sigma}^{O2}|,\qquad
|\hat{\sigma}^{N1}-\hat{\sigma}^{N2}| \gg |\hat{\sigma}^{N1}+\hat{\sigma}^{N2}|.  
\label{OONN}
\eeq
Since there is no reliable nonperturbative information on these functions,
it is natural to expect that all four functions $O(x,x)\pm O(x,0)$ and $N(x,x)\mp N(x,0)$
have a similar magnitude.  In this case, $A_N^D$ is mostly determined by
$O(x,x)+O(x,0)$ and $N(x,x)-N(x,0)$, while $O(x,x)-O(x,0)$ and $N(x,x)+N(x,0)$ can be neglected
by the relation (\ref{OONN}). 
From this observation, one obtains a modest
estimate on the upper bound for the combinations as 
\beq
&&|O(x,x)+O(x,0)| \leq (0.0003\sim 0.0004)xG(x),\nonumber\\
&&|N(x,x)-N(x,0)| \leq (0.0003\sim 0.0004)xG(x),  
\label{estimate}
\eeq
if $|A_N^D| \leq O(5\%)$.  
We remind that the relation (\ref{OONN}) is a peculiar feature for the process
$p^\uparrow p\to DX$ and does not generally hold in $ep^\uparrow \to eDX$\,\cite{BKTY}.

\section{Summary}
\vspace{1mm}

In this work, we have studied the contribution of the triple-gluon 
correlation functions to $A_N^D$ in $p^\uparrow p \to DX$.  
We derived the corresponding twist-3 single-spin-dependent cross section
in the leading order with respect to the QCD coupling constant.
The complete cross section receives contribution from
the four functions $O(x,x)$, $O(x,0)$, $N(x,x)$ and $N(x,0)$ as in the case 
of $ep^\uparrow \to eDX$\,\cite{BKTY},
and differs from the result in a 
previous work\,\cite{KQVY}.  
We have also presented a model calculation for $A_N^D$ at the RHIC energy. 
Assuming $|A_N^D|\leq O(5\%)$ as suggested by the RHIC preliminary data, we obtained a modest estimate
(\ref{estimate}) for a particular combination of the triple-gluon correlation functions.  
The detail of our analysis will be reported elsewhere\,\cite{KY}.

\section*{Acknowledgments}
We thank D. Boer, Z.-B. Kang, K. Tanaka, M. Liu, J.-W. Qiu and
F. Yuan for useful discussions, and the authors of Ref. \cite{KKKS} for providing us with
the fortran code of their $D$-meson fragmentation function.  
The work of S. Y. is supported by the Grand-in-Aid for Scientific Research
(No. 22.6032) from the Japan Society for the Promotion of Science.

\end{document}